\documentclass{article}
\usepackage{arxiv}
\usepackage[comma,authoryear]{natbib}
\usepackage[utf8]{inputenc} 
\usepackage[T1]{fontenc}    
\usepackage{hyperref}       
\usepackage{url}            
\usepackage{booktabs}       
\usepackage{amsmath,amsfonts,amssymb}       
\usepackage{nicefrac}       
\usepackage{microtype}      
\usepackage{lipsum}
\usepackage{algorithm}
\usepackage{algorithmic}
\usepackage{hyperref}
\usepackage{color}

\usepackage{amsmath,amsfonts,bm}

\def\eqref#1{equation~\ref{#1}}

\def\1{\bm{1}}

\DeclareMathAlphabet{\mathsfit}{\encodingdefault}{\sfdefault}{m}{sl}
\SetMathAlphabet{\mathsfit}{bold}{\encodingdefault}{\sfdefault}{bx}{n}

\usepackage{subcaption}
\usepackage{booktabs}

\usepackage{graphicx}
\graphicspath{ {./fig/} }

\title{COVID-19 on Social Media: Analyzing Misinformation in Twitter Conversations}

\author{
  Karishma Sharma,\hspace{0.1cm} Sungyong Seo,\hspace{0.1cm} Chuizheng Meng,\hspace{0.1cm} Sirisha Rambhatla,\hspace{0.1cm} Yan Liu \\
  Department of Computer Science\\
University of Southern California \\
Los Angeles, USA. \\
  \texttt{\{krsharma,sungyons,chuizhem,sirishar,yanliu.cs\}@usc.edu} \\
}

\begin{document}
\maketitle

\begin{abstract}
The ongoing Coronavirus (COVID-19) pandemic highlights the inter-connectedness of
our present-day globalized world. With social distancing policies in place, virtual communication
has become an important source of (mis)information. As increasing number of people rely on
social media platforms for news, identifying misinformation and uncovering the nature of online discourse around COVID-19 has emerged as a critical task.  To this end, we collected streaming data related to COVID-19 using the Twitter API, starting March 1, 2020. We identified unreliable and misleading contents based on fact-checking sources, and examined the narratives promoted in misinformation tweets, along with the distribution of engagements with these tweets. In addition, we provide examples of the spreading patterns of prominent misinformation tweets. The analysis is presented and updated on a publically accessible dashboard\footnote{\url{https://usc-melady.github.io/COVID-19-Tweet-Analysis}} to track the nature of online discourse and misinformation about COVID-19 on Twitter from March 1 - June 5, 2020. The dashboard provides a daily list of identified misinformation tweets, along with topics, sentiments, and emerging trends in the COVID-19 Twitter discourse. The dashboard is provided to improve visibility into the nature and quality of information shared online, and provide real-time access to insights and information extracted from the dataset.


\keywords{COVID-19 \and Misinformation \and Fake News \and Social Media}
\end{abstract}

\section{Introduction}
\label{intro}

Social media plays a pivotal role in information dissemination and consumption during a pandemic;  more so with increasing social distancing and growing reliance on online communication \citep{SocialMediaCovidRole}. It has both positive and negative social impacts during the crisis. For instance, safety tips such as ``wash your hands" and  ``stay home" are shared widely to gain community support in fighting the COVID-19 pandemic \citep{SocialMediaCovidRole}. On the other hand, misinformation and hate speech are growing problems that can adversely impact the safety of individuals and society. The associated public health risk and other dire consequences of misinformation spread on social media - such as 5G towers being burned down due to conspiracy theories linking them to COVID-19 \citep{5GCOVID}, make it imperative to address the problem of misinformation.

In recent years, the reliance on social media as a source of news has continued to increase \citep{Mitchell2018,Geiger2019}. Here, we focus our analysis on the social media platform, Twitter, to understand the nature of online discourse surrounding COVID-19, since Twitter has the highest number of news focused users \citep{PewTwitter2019}. The impact of misinformation surrounding the COVID-19 pandemic can be especially damaging, since any missteps can increase the chances of an exponential spread of the disease, or accidental death due to self-medication \cite{NYTArizona2020}. The emergence of COVID-19 related misinformation on online platforms motivated the World Health Organization (WHO) to launch a ``Mythbuster'' page \cite{WHO2020} at the start of the pandemic. However, such counter measures are limited in their ability to combat misinformation, due to the large-scale nature and fast-paced evolution of online discourse. 

Therefore, we utilize computational solutions to extract and examine the nature of social media conversations surrounding COVID-19. We  collected  streaming data related to the pandemic using the Twitter API, starting March 1, 2020. We identified unreliable and misleading contents based on fact-checking sources, and examined the narratives promoted in misinformation tweets, along with the distribution of engagements with these tweets. In addition, we provide examples of the spreading patterns of prominent misinformation tweets. 

\begin{figure}[t]
\begin{subfigure}{.5\textwidth}
  \centering
  \includegraphics[width=\linewidth, height=5cm]{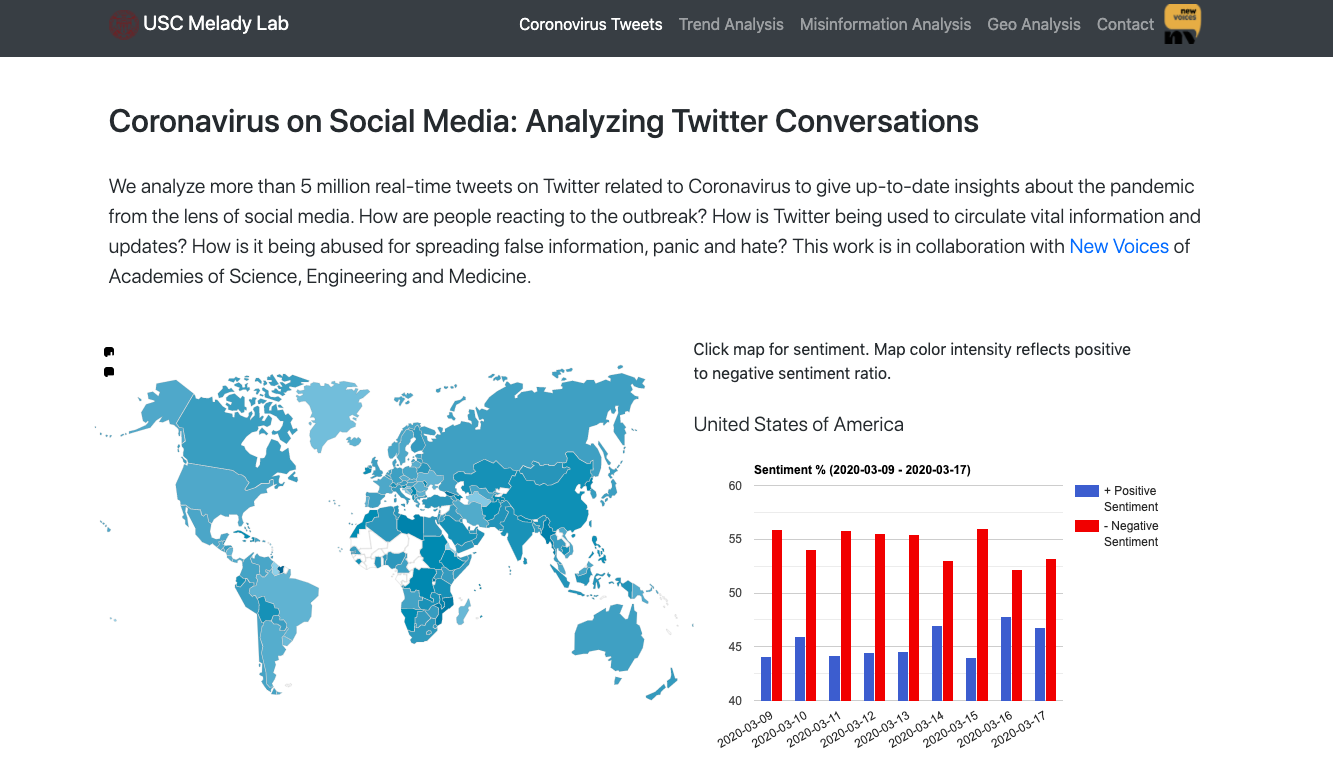}  
  \caption{Sentiment and topic analysis of Twitter conversations.}
  \label{fig:sub-first}
\end{subfigure}
\begin{subfigure}{.5\textwidth}
  \centering
  \includegraphics[width=\linewidth, height=5cm]{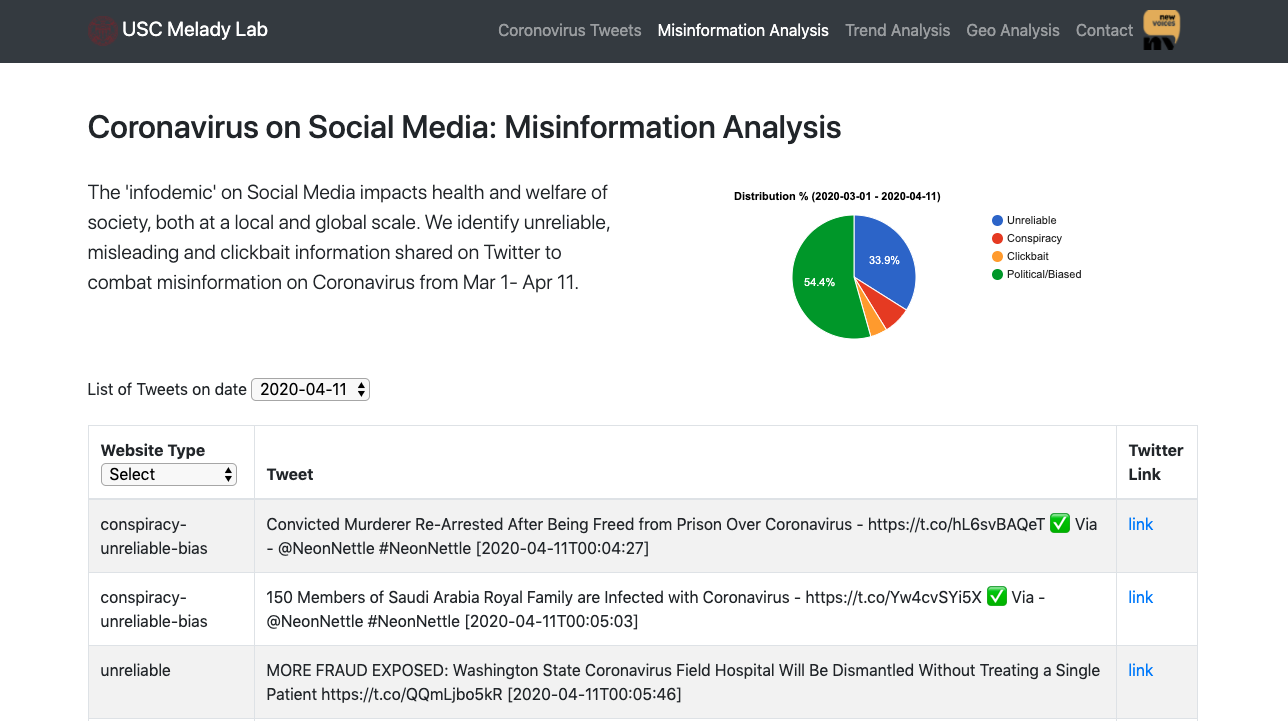}  
  \caption{Misinformation analysis on Twitter}
  \label{fig:sub-second}
\end{subfigure}
\caption{COVID-19 Social Media Analysis Dashboard analyzing sentiments, trends, and misinformation on Twitter. Accessible online at \url{https://usc-melady.github.io/COVID-19-Tweet-Analysis}.}
\label{fig:dashboard}
\end{figure}

In order to provide timely insights about the narratives and quality of information shared on social media during the pandemic, the analysis is presented as a publically accessible dashboard (Fig~\ref{fig:dashboard}) - tracking the online Twitter discourse on COVID-19 from March 1 - June 5, 2020. The dashboard provides a daily list of identified misinformation tweets, along with topics, sentiments, and emerging trends in the COVID-19 Twitter discourse. The dashboard is provided to improve visibility into the nature and quality of information shared, and provide real-time access to insights and information extracted from the dataset. We aim to improve public awareness of the broader nature of online discourse related to the pandemic, towards enabling better discernment of misinformation claims.

\section{Related Work}
\label{relwork}

The first outbreak of the COVID-19 pandemic (Novel Coronavirus Disease) was reported in Wuhan, China in late December, 2019; and has rapidly spread to  several  countries  across  the  world during  early  March,  with  the  United States declaring a national emergency on March 13, 2020 [11]. The rapid infection rates and enforcement of social distancing policies quickly promoted widespread  online  discussions  regarding  the  pandemic  on  social  media  plat-forms; with a growing number of conspiracies and false information.

Misinformation on social media in general has been an increasingly pertinent problem in recent years \citep{sharma2019combating}. The proliferation of misinformation related to elections, hurricanes, earthquakes, and civil discourse has been studied in different social and political contexts \citep{allcott2017social,gupta2013faking}. With the COVID-19 pandemic, there have been several recent works analyzing different aspects of misinformation on social media. The majority of these studies are pre-print papers (accessed June, 2020) with early characterization of social media discussions on the topic \citep{cinelli2020covid,cui2020coaid,singh2020first}. 

The recent work in \cite{singh2020first} analyzed the temporal correlation between social media conversations on Twitter and disease outbreaks at different locations; and considered the distribution of information from different sources, between January and March, 2020. \cite{cui2020coaid,zhou2020recovery} created a dataset based on fact-checked claims related to COVID-19 for benchmarking different automated detection models \citep{ruchansky2017csi,shu2019defend}. \cite{cinelli2020covid} modeled the spread of information on social media using epidemic models used in disease spread modeling \cite{newman2002spread} for different social media platforms. 


In analysis of social bots during the pandemic, \cite{ferrara2020types} provided an estimation and characterization of user accounts that reflected automation or bot-like characteristics in COVID-19 discussions on social media. Their findings provide early evidence of the existence of automated accounts used as  news-posting bots, and ones promoting political conspiracies. Similarly, \cite{huang2020disinformation} provided analysis of different types of user accounts. The analysis of bots and coordinated influence campaigns have been an area of active research in the context of Russian interference in the 2016 US elections \citep{bessi2016social,luceri2020detecting,pacheco2020uncovering,sharma2020identifying}. 

Other studies aim to understand the psychological impacts of misinformation on public perception, and sharing of false information on social media \citep{pennycook2020fighting,swami2020analytic}. \cite{pennycook2018falls} 
conducted a study with 1,600 participants, and found that
participants were far less discerning when deciding to share true or false information on social media, relative to when asked directly about correctness of the information. In another study of 520 subjects, \cite{swami2020analytic} identified that analytic thinking and rejection of COVID-19 conspiracy theories, respectively, were significantly and directly associated with compliance to mandated social-distancing measures. These results have important implications regarding the risks of misinformation on public perception and in turn, on the effectiveness of health intervention policies aimed at controlling the spread of the pandemic.


Here, we address the broader nature of the online discourse surrounding the COVID-19 pandemic, through examination of misinformation claims, and extraction of topics, sentiments and trends from Twitter conversations. The daily analysis of misinformation claims and online discourse from March 1 to June 5, 2020 on the data collected through the Twitter streaming API was made publically available through the dashboard, with the aim of providing real-time insights and analysis of the online discourse, towards enabling better public discernment of misinformation claims about COVID-19.

To tackle the real-time surge of misinformation on social media related to the COVID-19 pandemic, traditional methods to annotate each claim are not scalable with the magnitude of global conversations surrounding the pandemic. Therefore, since earlier studies have shown strong correlation between credibility of news and news publishing websites; with evidence that most fake news is generated from low credibility or hyperpartisan websites \citep{zhou2018fake}, we obtain and leverage information about news publishing websites from fact-checking sources to identify misinformation.

\section{Data Collection}
We collected the dataset using the Twitter streaming API service\footnote{\url{https://developer.twitter.com/en/docs/tweets/filter-realtime/guides/basic-stream-parameters}} from March 1, 2020 to June 5, 2020. We used keywords related to COVID-19 to filter the Twitter stream and obtain relevant tweets about the pandemic. The stream fetches a 1\%  random sample of all tweets containing at least one of the keywords (`Covid19', `coronavirus', `corona virus', `2019nCoV ', `CoronavirusOutbreak', `coronapocalypse'), from the platform in real time. The COVID-19 crisis was declared a global pandemic on March 11, 2020, which motivated the mentioned collection period. From March 1-8 and March 18-19, due to interruptions in fetching the stream, we recovered data using tweet ids in \cite{chen2020tracking}.

\textbf{Dataset statistics.}  Table~\ref{tab:dataset}  provides details about the tweets collected and the user accounts associated with the tweets. The dataset contains 85.04M tweets from 182 countries. The subset of English tweets equals 54.32M. The English tweets are utilized for further analysis and therefore the table reports the details about what fraction of English tweets contains geolocation information, and count of unique user accounts associated with the tweets, as well as the percentage of Twitter verified accounts among those user accounts. 

\textbf{Geolocation.} The tweet geolocation information at the country-level is extracted directly from tweet metadata if it has geo-location enabled, otherwise extracted from the  user reported locations in the user profile information associated with the tweet \citep{dredze2013carmen}. The geo-location information is not always be available, in cases when the user reported location is not a valid geographical location. Geo-location enabled tweets are sparse, therefore geo-location from user profiles is included. Table~\ref{tab:geographical_dist} provides the distribution of English tweets across countries, and the distribution of English tweets across US states.

\begin{table}[t]
\centering
\caption{Statistics of COVID-19 tweets collected between March 1-June 5, 2020.}
\label{tab:dataset}
 \begin{tabular}{l|r}
\toprule
\textbf{Dataset(March 1 - June 5, 2020)} & \textbf{Statistics}  \\
\midrule
\# Tweets & 85.04 M \\
\% Tweets (In English Lang) & 63.88\%  \\
\% Tweets (with Geo Information) &  43.02\% \\
\# User Accounts & 10.61 M \\
\% Verified User Accounts & 7.51\% \\
\bottomrule
\end{tabular}
\end{table}

\begin{table}[t]
\renewcommand{\arraystretch}{1.1}%
\centering
\caption{The distribution of English tweets across countries and across US states.}
\begin{tabular}{l|r} 
\toprule
\textbf{Top Countries} & \textbf{\# Tweets} \\
\midrule
United States        & 13290691 \\
United Kingdom       & 2657375  \\
India                & 1829656  \\
Canada               & 884420   \\
Nigeria              & 698603   \\
Australia            & 510069   \\
South Africa         & 278457   \\
Pakistan             & 223659   \\
Kenya                & 201537   \\
Ireland              & 188229   \\
France               & 162343   \\
Malaysia             & 151586   \\
Hong Kong            & 126006   \\
Philippines          & 122805   \\
Indonesia            & 113489   \\
Germany              & 109109   \\
Ghana                & 94975    \\
Uganda               & 70814    \\
Spain                & 68993    \\
UAE                  & 67224    \\
\bottomrule
\end{tabular}
\qquad
\begin{tabular}{l|r}
\toprule
\textbf{Top US States} & \textbf{\# Tweets} \\
\midrule
California           & 1958121 \\
New York             & 1091428 \\
Florida              & 906426  \\
Texas                & 883169  \\
Pennsylvania         & 437928  \\
Dist. of Columbia & 402157  \\
Illinois             & 387298  \\
Georgia              & 364500  \\
Ohio                 & 353748  \\
Arizona              & 311748  \\
Michigan             & 298246  \\
New Jersey           & 287030  \\
Virginia             & 284819  \\
Massachusetts        & 263473  \\
Washington           & 232426  \\
Oregon               & 221913  \\
North Carolina       & 219956  \\
Maryland             & 217763  \\
Tennessee            & 215048  \\
Colorado             & 177931  \\
\bottomrule
\end{tabular}
\label{tab:geographical_dist}
\end{table}


\section{Misinformation}
Increased reliance on social media for news, and the risk of misinformation exposure on public health, have made tackling of misinformation claims time critical. Existing misinformation datasets are either pertaining to general newsworthy events during a particular time period \citep{ma2016detecting,ma2018detect}, or domain specific, such as related to the Syrian civil war or Hurricane Sandy \citep{salem2019fa,gupta2013faking}. In this work, we create a domain specific dataset on COVID-19 leveraging knowledge from fact-checking sources, for analysis of misinformation related to the pandemic, and in the future, to improve research in misinformation related to healthcare and pandemics.


\textbf{Categorization of news sources}  The credibility of news is positively correlated with the credibility of the news publishing websites/sources publishing the news. Most fake news has been witnessed to be generated from low credibility and hyperpartisan websites on social media \citep{zhou2018fake}. Fact-checking sources (eg. Snopes, PolitiFact, Media Bias/Fact check, NewsGuard) conduct independent journalistic verification on the credibility of both, individual claims surfaced on social media, as well as the associated news publishing websites linked to false, unreliable and misleading claims. We compile information from three  fact-checking sources that provide extensive journalistic analysis of low-quality news sources known to frequently publish unreliable and false information. The three fact-checking sources considered in this work are Media Bias/Fact Check\footnote{\url{https://mediabiasfactcheck.com/}}, NewsGuard\footnote{\url{https://www.newsguardtech.com/covid-19-resources/}} and \cite{zimdars2016false}\footnote{\url{https://docs.google.com/document/d/1zhaZooMfcJvk_23in201nviWJN1-LhRvGlPXJWBrPRY/edit?usp=sharing}}. 

NewsGuard maintains a repository of news publishing sources that have published false information about COVID-19. We include Media Bias/Fact Check's list of questionable news sources with low and very low factual reporting. \cite{zimdars2016false} maintains a list of different kinds of false, conspiracy, junk-science, clickbait and other types of news sources.
Note that a tweet can belong to more than one type; as can multiple types be associated with a news source. We assign the news sources to the above mentioned types based on the following criteria. For Media Bias/Fact Check, we include the list of questionable news sources with reported low and very low factual content, into the unreliable categorization. Similarly, we include news sources listed by NewsGuard for publishing false content related to COVID-19 into the unreliable categorization. In the case of \cite{zimdars2016false}, we include tags fake, rumor, unreliable, and satire, in the unreliable categorization. We include tags conspiracy and junksci (pseudoscience, naturalistic fallacies) in the conspiracy categorization; clickbait tag in the clickbait categorization; and tags bias and political to the political/biased categorization. The definitions of the different tags can be found in \cite{zimdars2016false}\footnote{\url{https://docs.google.com/document/d/1zhaZooMfcJvk_23in201nviWJN1-LhRvGlPXJWBrPRY/edit?usp=sharing}}. As in \cite{zimdars2016false}, if a news source is associated with more than one tag, we include it in the each of the categories associated with its tags. For instance, `political-clickbait' is an example of a news source publishing misleading information of a politically biased and clickbait type. We exclude news sources with solely the political tag in \cite{zimdars2016false} and none of the other tags associated with the other categorizations, as it consists of reliable news with a bias, but is not false or misleading.

We categorize information into four types - \emph{unreliable}, \emph{conspiracy}, \emph{clickbait}, and \emph{political/biased}. News sources can be categorized into multiple types, and therefore, the associated tweets can have more than one type label.

\begin{itemize}
    \item \emph{Unreliable}. We define the unreliable category to include false, questionable, rumorous and unreliable news. In addition, we include satire, based on the consideration that satire has the potential to perpetuate misinformation \citep{zimdars2016false} or be used as a cover for misinformation publication \citep{sharma2019combating}.
    
    \item \emph{Conspiracy.} We define conspiracy to include conspiracy theories and scientifically dubious news.

    \item \emph{Clickbait.} This category includes clickbait news i.e. misleading, distorted or exaggerated headlines and/or body purposed to attract attention, for reliable and/or unreliable information.
    
    \item \emph{Political/Biased.} This category includes political and biased news, written in support of a particular point of view or political orientation, for reliable and/or unreliable information such as propaganda.
\end{itemize}

\textbf{Extraction of information cascades} Information cascades \citep{yang2010modeling} represent the diffusion or spread of information over social media. They can be considered as a time ordered sequence of posts originating from a source post and spreading through re-shares or engagements as retweets and replies to the source and subsequent posts. 

The retweet/reply graph contained 42.71M edges i.e. retweet or reply links between the 54.32M English tweets collected in the dataset. The retweet or reply links are extracted from the tweet metadata collected using the Twitter API. We extract weakly connected components of the retweet/reply graph to identify source post cascades being propagated or shared over the social network. Each weakly connected component is a directed tree rooted at the source post, with other vertices and edges in the tree representing retweets and replies of the source or subsequent posts in the tree. 

Each source tweet has associated tweet text and user account features, and the extracted cascade originating at the source post. We label the source tweet as misinformation (unreliable, conspiracy, clickbait, political/biased) if the tweet shared any article or content posted from any of the misinformation sources compiled using the fact-checking sources. The source tweet metadata is used to identify external links to news articles referenced by the tweet.

\section{Misinformation Analysis}

\begin{table}[t]
    \centering
    \caption{Dataset statistics of source tweets in misinformation cascades}
    \begin{tabular}{c|c|c|c}
        \toprule
         Tweets (EN) & Source Tweets & Source Tweets (URLs) & Misinformation Source Tweets \\
         \midrule
         54.32M & 6.37M & 4.58M & 150.8K \\
         \bottomrule
    \end{tabular}
    \label{tab:stats_misinfo_source_tweet}
\end{table}


\begin{figure}[t]
    \centering
    \includegraphics[scale=0.4]{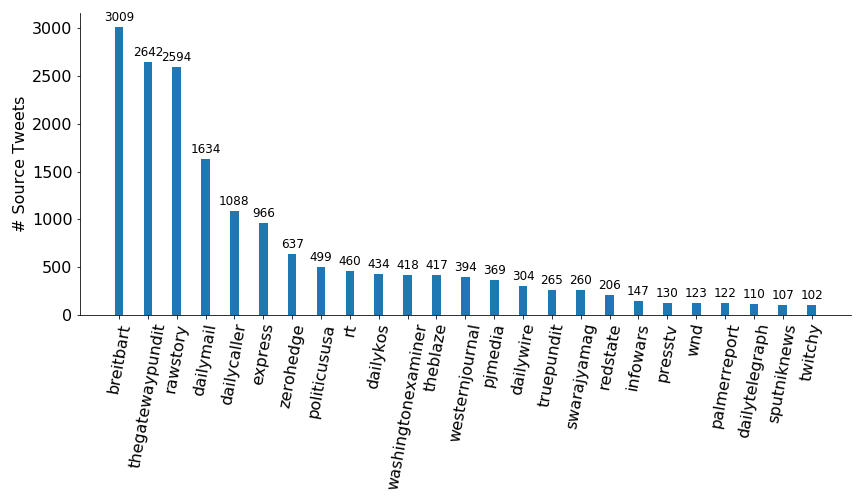}
    \caption{Distribution of news publishing sources linked to misinformation tweets}
    \label{fig:breakdown}
\end{figure}

\begin{figure}[t]
    \centering
    \includegraphics[scale=0.4]{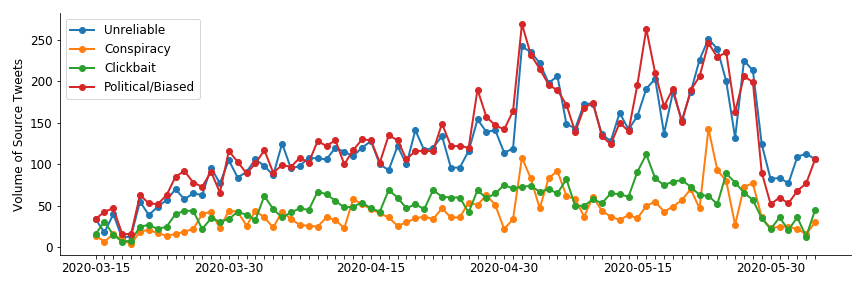}
    \caption{Volume of misinformation source tweets by type per day in March-June}
    \label{fig:volume}
\end{figure}


In this section, we analyze the distribution of misinformation tweets, and distribution of engagements with the misinformation tweets from the collected dataset. This is followed by identifying textual narratives in the tweets though distinctive hashtags promoted in the misinformation tweets. Lastly, we provide examples of misinformation cascades and visualize their propagation based on extracted geo-location information of tweets in the cascades.

\subsection{Misinformation dataset statistics} 
In this section, we first analyze the distribution of the misinformation cascades identified in the datasets using information from fact-checking sources, as detailed in the previous section. Table~\ref{tab:stats_misinfo_source_tweet} provides the statistics of identified misinformation tweets. As seen, 150.8K (3.29\%) of source tweets with external links (urls) constitute the misinformation source tweets that link to the unreliable, conspiracy, clickbait, political/biased news sources.

We further explore the distribution of low quality news publishing sources that are most widely linked to source tweets in misinformation cascades. In Fig~\ref{fig:breakdown} the breakdown of misinformation source tweets according to news publishing sources (websites) is provided. The figure shows the top 25 most frequently linked news sources in source tweets of the dataset. The counts of source tweets that link to each source are provided. The top 25 news sources were linked to more than a 100 source tweets in the collected dataset. 

\subsection{Misinformation narratives}

We leverage the misinformation categorization to examine the different narratives spread from low quality news publishing sources, on social media. We analyze the distribution of engagements in misinformation cascades, followed by textual analysis of the narratives of each type of misinformation.

\begin{figure}[t]
    \centering
    \includegraphics[scale=0.4]{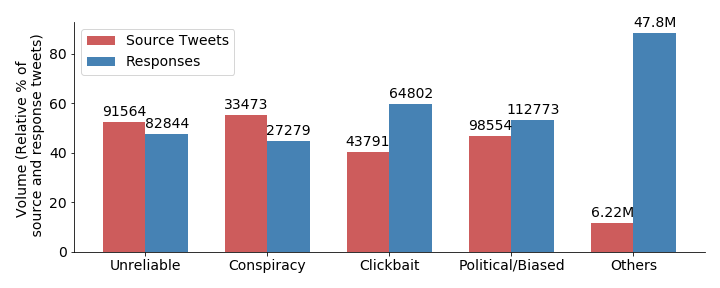}
    \caption{Relative volume of source tweets to responses (replies/retweets) for each misinformation category.}
    \label{fig:relative_volume}
\end{figure}

\begin{table}[t]
\caption{Top hashtags in misinformation category ordered by TF-IDF scores}
\resizebox{0.98\textwidth}{!}{
\begin{tabular}{llll}
\toprule
Unreliable & Conspiracy & Clickbait & Political/Biased \\
\midrule
vaccine               & wwg1wga            & trumpvirus               & kag                 \\
scummedia             & wuhanvirus         & coronaviruspandemic      & wwg1wga             \\
wuhanvirus            & freezerohedge      & trumpgenocide            & fakenews            \\
ibackboris            & lockdown           & unitedstates             & wuhanvirus          \\
lockdown              & billgates          & pandemic                 & trump2020           \\
kag                   & fakenews           & commondreams             & pandemic            \\
wwg1wga               & chinavirus         & politics                 & feedly              \\
fakenews              & pandemic           & gop                      & chinavirus          \\
covid19news           & hydroxychloroquine & foxnews                  & coronaviruspandemic \\
covid19india          & kag                & trumpisanationaldisgrace & chinesevirus        \\
indonesiabebascovid19 & who                & usa                      & kag2020             \\
fightingcovid19       & infowars           & kag                      & democrats           \\
covid19testing        & ats                & trumpliespeopledie       & lockdown            \\
covid19memes          & firefauci          & tcot                     & tcot                \\
coronaviruscovid19    & ccpvirus           & trumpownseverydeath      & politics            \\
bajucovid19           & vaccine            & trumpliesamericansdie    & hydroxychloroquine  \\
chinavirus            & fauci              & usrc                     & usa                 \\
ibackdom              & plandemic          & resistance               & arrestthemallnow    \\
usnews                & corona             & fakenews                 & ricothedems         \\
uk                    & feedly             & moscowmitch              & impeachobamasjudges \\
\bottomrule
\end{tabular}
}
\label{tab:narratives}
\end{table}

In Fig~\ref{fig:volume}, we examine the  volume and distribution of source tweets over time for each category of misinformation, per day during the observed time period from March to June as shown. The volume of source tweets of misinformation cascades for each category (unreliable, conspiracy, clickbait, political/biased) are shown separately. In the case of tweets that belong to more than one category, we count the tweet in each of the categories associated with it. The volume of misinformation source tweets increases with time, as does the volume of overall tweets in the dataset, from March to June. This increase can be indicative of an increase in the global activity and online discussions around COVID-19 due to spread of the pandemic to multiple countries starting from early March. The volume of source tweets linked to conspiracy and clickbait sources is smaller compared the unreliable and political/biased sources in the collected social media and news sources data.

In Fig~\ref{fig:relative_volume}, we explore the distribution of engagements in each category from the extracted misinformation cascades. The figure shows the relative distribution of source tweets to responses (replies/retweets of the source tweets) in misinformation cascades for each category. As seen, the Unreliable and Conspiracy types receive fewer responses relative to the volume of source tweets, as compared to other categories. The ``Others" category constitutes information cascades which are not labeled as misinformation as the source tweets in these cascades do not belong to any of the four types (unreliable, conspiracy, clickbait, political/biased). The Clickbait category distribution is closest to Others and Clickbait, Political/Biased receive more responses to source tweets. 

This finding suggests that false information might be harder to spread through general users on social media, as source tweets linked to low quality news are observed as less likely to be shared or replied to than others.

Lastly, we provide textual analysis of the source tweets in misinformation cascades by type. For textual analysis, we extract hashtags from the textual content of source tweets in identified misinformation cascades. We find distinctive hashtags in source tweets of each category, by computing the TF-IDF scores of hashtags across each category. The top distinctive hashtags are the ones with the highest TF-IDF scores, that do not appear in the top ten hashtags of the other three categories when ordered by the scores. The distinctive narratives in each category are shown in Table~\ref{tab:narratives}. The list of misinformation tweets is presented through the dashboard as a daily list, along with the category labels for public access to the misinformation claims circulated online during the pandemic.

\subsection{Misinformation spread across countries}

\begin{figure}[t]

\begin{subfigure}{.5\textwidth}
  \centering
  \includegraphics[width=\linewidth]{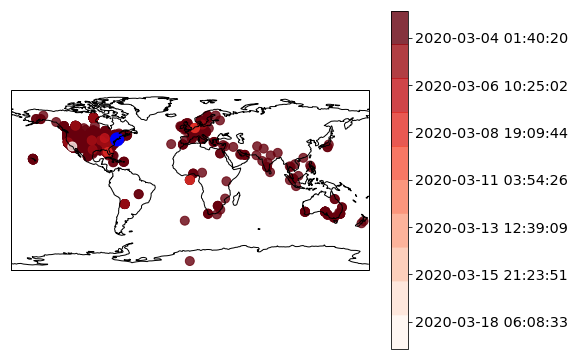}  
  \caption{\textbf{Political-Clickbait.} ``GOP blocking coronavirus bill — because it limits how much drugmakers can charge for a vaccine."}
  \label{spread_clickbait}
\end{subfigure}
~
\begin{subfigure}{.5\textwidth}
  \centering
  \includegraphics[width=\linewidth]{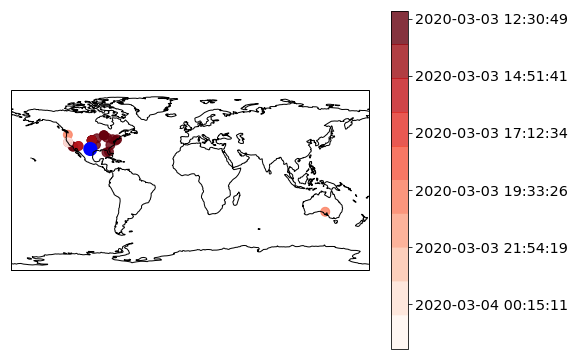}  
  \caption{\textbf{Unreliable.} ``An Obama Holdover in an Obscure Government Arm Helped Cause the Country’s Coronavirus Crisis."}
  \label{spread_obama}
\end{subfigure}
\caption{Misinformation spread across, political examples. (Left)  Example with political-clickbait news (Right) Example with unreliable news. Legend: Source tweet (Blue), Retweet/Replies (Red, intensity based on time scale).}
\label{fig:misinfo_spread}
\end{figure}

The misinformation spread across countries for sample tweets identified from the collected dataset is shown in Figure~\ref{fig:misinfo_spread}. The figure shows the information cascade corresponding to each source tweet. The points indicate the retweet or replies of the source tweet over the time scale. Tweets containing geolocation information are visualized, based on the extracted latitude and longitude information. The identified misinformation in the four categories - unreliable, conspiracy, clickbait, and political/biased, were found to contain both healthcare and political misinformation. We provide and discuss examples of source tweets and their propagation patterns across countries.

\begin{figure}[t]

\begin{subfigure}{0.5\textwidth}
  \centering
  \includegraphics[width=\linewidth]{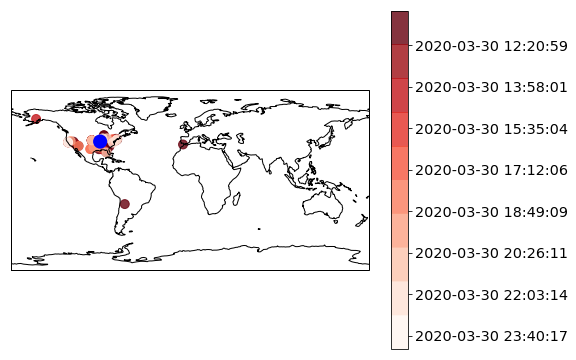}  
  \caption{\textbf{Unreliable.} ``Nevada Governor Sisolak's Chief Medical Officer Who Banned Hydroxychloroquine for Treating Coronavirus DOES NOT Have License to Practice Medicine."}
  \label{spread_hydroxy}
\end{subfigure}
~
\begin{subfigure}{0.5\textwidth}
  \centering
  \includegraphics[width=\linewidth]{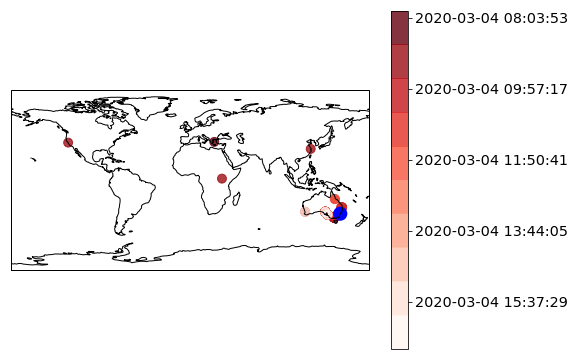}  
  \caption{\textbf{Unreliable.} ``Widespread panic hits as Corona Virus found to transmit via toilet paper \#toiletpapergate \#coronavirus"}
  \label{spread_toiletpaper}
\end{subfigure}
\caption{Misinformation spread across countries. (Left) Source tweet observed in United States (Right) Source tweet observed outside of United States. Legend: Source tweet (Blue), Retweet/Replies (Red, intensity based on time scale).}
\end{figure}

\textbf{Discussion.} In Fig.~\ref{spread_hydroxy}, a  false claim circulated about Nevada Governor's Chief Medical Officer banning the use of Hydroxychloroquine treatments was seen to circulate through social media. In this case, the observed geolocation of source tweet is in the United States, the country with the highest Twitter usage, and it propagates to other countries within minutes. In other cases, source tweets are also observed to originate from other countries and travel to United States and other countries. For example, in Fig.~\ref{spread_toiletpaper}, for a false claim that the virus was found to transmit through toilet paper, the source tweet geolocation was observed in Australia, with retweets traveling to several other countries.

\begin{figure}[t]

\begin{subfigure}{.5\textwidth}
  \includegraphics[width=\linewidth]{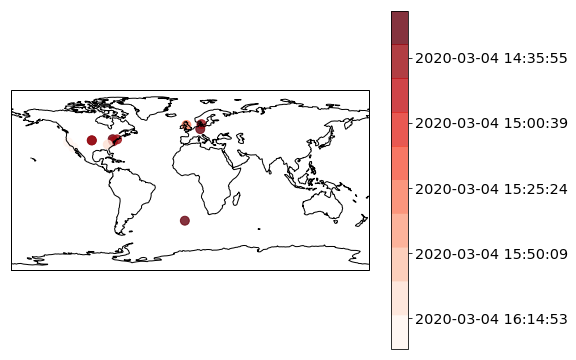}  
  \caption{\textbf{Conspiracy.} ``It is not the flu, it is a \texttt{\#bioweapon}."}
  \label{spread_bioweapon}
\end{subfigure}
~
\begin{subfigure}{.5\textwidth}
  \centering
  \includegraphics[width=\linewidth]{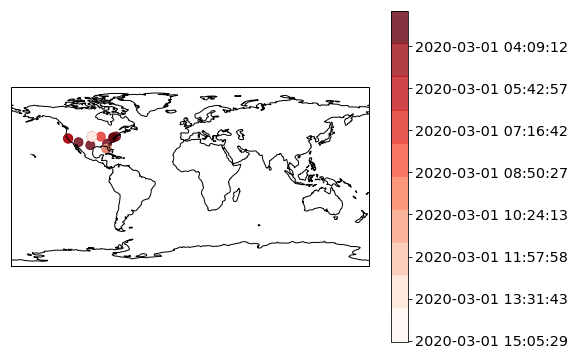}  
  \caption{\textbf{Unreliable.} ``Numbers Show Coronavirus Appears Far Less Deadly Than Flu Media Keep Promoting Panic."}
  \label{spread_panic}
\end{subfigure}

\caption{Misinformation spread across countries, examples with source tweet geolocation unavailable. (Left) Retweets/Replies observed across countries (Right) Retweets/Replies observed within United States. Legend: Retweet/Replies (Red, intensity based on time scale).}
\end{figure}

In cases where the geolocation information of the source tweet is unobserved, the geolocation of other tweets in the information cascade still provides estimates of the exposure and spread in different countries, of the misinformation claims propagated through the source tweet. For instance, in Fig.~\ref{spread_bioweapon} the spread of the conspiracy promoting that the virus is a bioweapon is observed over multiple countries, whereas in Fig.~\ref{spread_panic} the spread of the claim that the 
pandemic is less deadly than the flu is observed within the United States.

Misinformation of varying degree of falsehood and biased/clickbait news reporting can mislead and influence public perception, especially with widespread propagation. We observe that the largest 
cascade in the collected dataset has over $10,000$ retweets spanning multiple countries, shown in Fig~\ref{spread_clickbait}. It corresponds to a political clickbait news article published on the discussion surrounding affordability and price control on vaccines being researched for the virus. We also find other cases of political misinformation with false claims regarding political figures maliciously attempting to worsen the crisis, as shown in Fig.~\ref{spread_obama}. As seen these cases of misinformation have the potential to harm public health and effectiveness of health intervention policies.


\label{sentiment_topic_trend}

\section{Sentiment Analysis}

\begin{figure}[t]

\begin{subfigure}{\textwidth}
   \includegraphics[width=\linewidth, height=5cm]{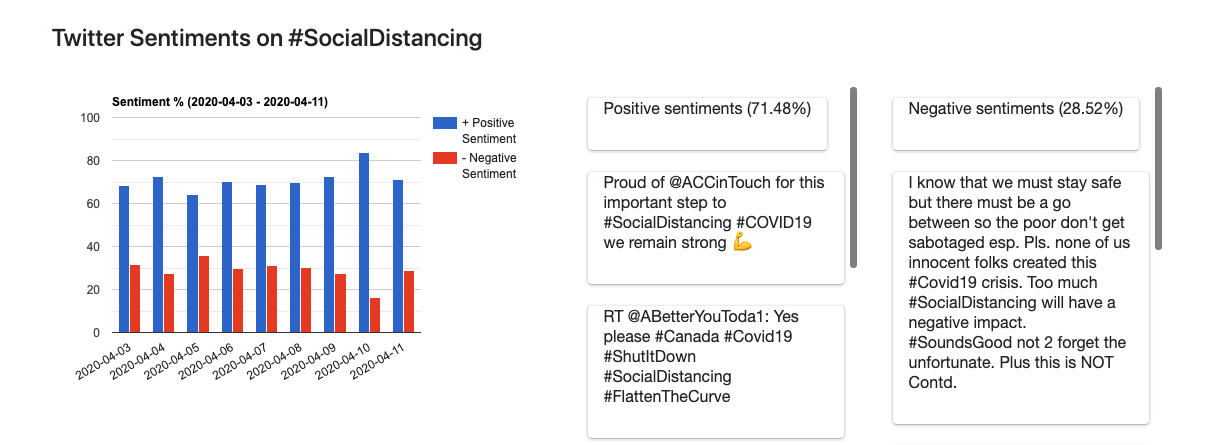}  
  \label{fig:sentiment_sd}
\end{subfigure}
~
\begin{subfigure}{\textwidth}
  \centering
  \includegraphics[width=\linewidth, height=5cm]{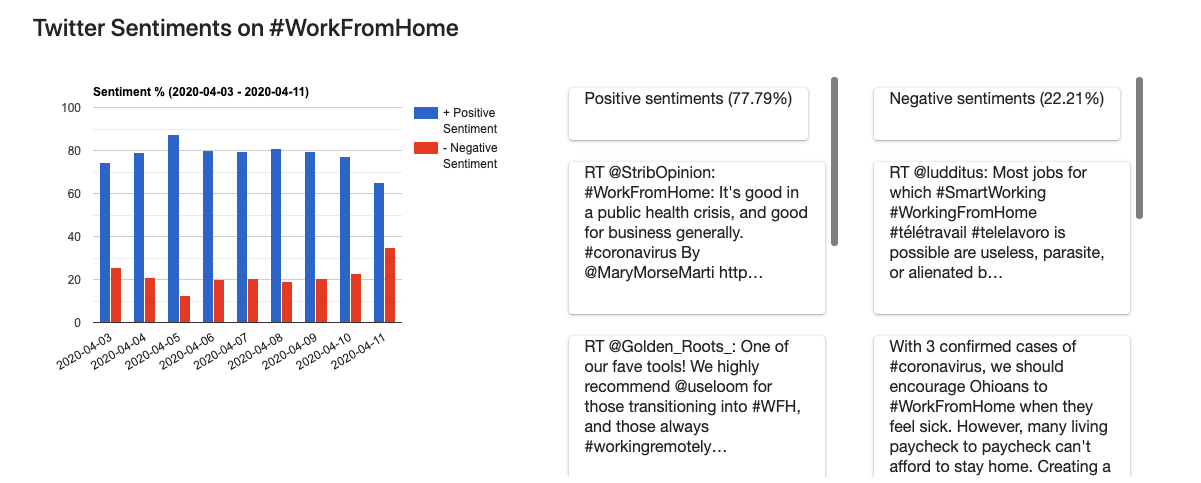}  
  \label{fig:sentiment_wfh}
\end{subfigure}
\caption{Twitter sentiments analysis related to intervention policies of SocialDistancing and WorkFromHome.}
\label{fig:sentiment_policies}
\end{figure}

\textbf{Country-wise sentiments.} We analyze the evolving country-wise sentiments related to the COVID-19 pandemic. The public perceptions
constitute an important factor for gauging the reactions to policy decisions and preparedness efforts. In addition, they
also reflect the nature of news coverage and potential misinformation. We extract sentiments from social media posts at
the country-level and over time, to study the evolving public perceptions towards the pandemic. Using lexical sentiment
extraction based on  \citep{hutto2014vader}, we obtain the valence (positive or negative) along with its intensity for
each tweet based on its textual information. The sentiment is aggregated over tweets to estimate the overall sentiment
distribution. The distribution of sentiments was found to vary over time and country.

\textbf{Social distancing/Work from home sentiments.} In addition, we analyze the public perception of emerging policies such as social distancing and remote work. These
disease mitigation strategies also provide unprecedented glimpse into the effect of remote work and isolation on mental
health. Although the option to work remotely is limited to the white collar workforce, nevertheless absence of child and
dependent-care has emerged as an important challenge. Furthermore, this forced remote work will impact workdays
of white collar workers beyond the pandemic. In order to understand public sentiment and opinion about different
social issues, we extract hashtag information from the collected tweets, and filter based on keywords ``\texttt{\#workfromhome,
\#wfm, \#workfromhome, \#workingfromhome, \#wfhlife}'' and ``\texttt{\#socialdistance, \#socialdistancing}''. The filtered tweets
are analyzed to obtain positive and negative sentiments and ranked and visualized based on valence and intensity. The analysis is shown in Fig.~\ref{fig:sentiment_policies} for sentiments on social distancing and on work from home policy interventions.


\section{Topic and Trend Analysis}

\textbf{Topic clusters.} We analyze Twitter conversations to identify topics and trends in the Twitter data on COVID-19. We use topic modeling
based on character embeddings \citep{joulin2016fasttext} extracted from social media posts \cite{nguyen2015improving,li2016topic}. We identified 20 different topics from the collected English tweets. We found that the prominent topics of discussions during early
March were centered around global outbreaks (Wuhan, Italy, Iran), travel restrictions, prevention measures such as
hand washing and masks, hoarding, symptoms and infections, immunization, event cancellations, testing kits and
centers, government response and emergency funding. The topic clusters along with the most representative tweets in each cluster are provided on the dashboard. The representative tweets of each cluster are obtained based on word similarity of the tweet to the tf-idf word distribution of the cluster. The label to each cluster of tweets was assigned by manual inspection of the word distribution and representative tweets of the cluster.

\begin{figure}[t]

\begin{subfigure}{.5\textwidth}
  \centering
  \includegraphics[width=\linewidth,,height=4cm]{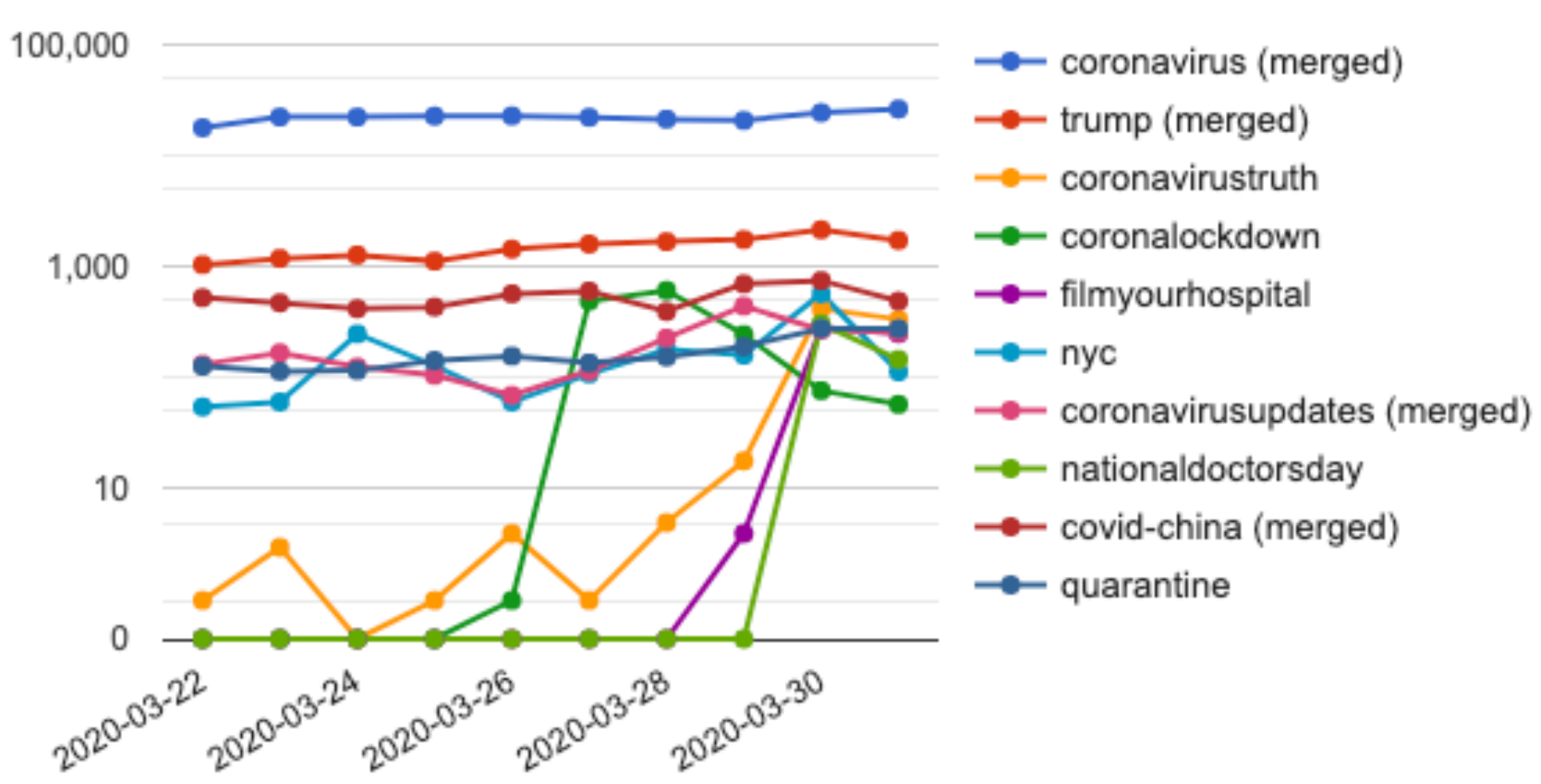}  
  \caption{United States, March 22-31.}
\end{subfigure}
~
\begin{subfigure}{.5\textwidth}
  \centering
  \includegraphics[width=\linewidth,height=4cm]{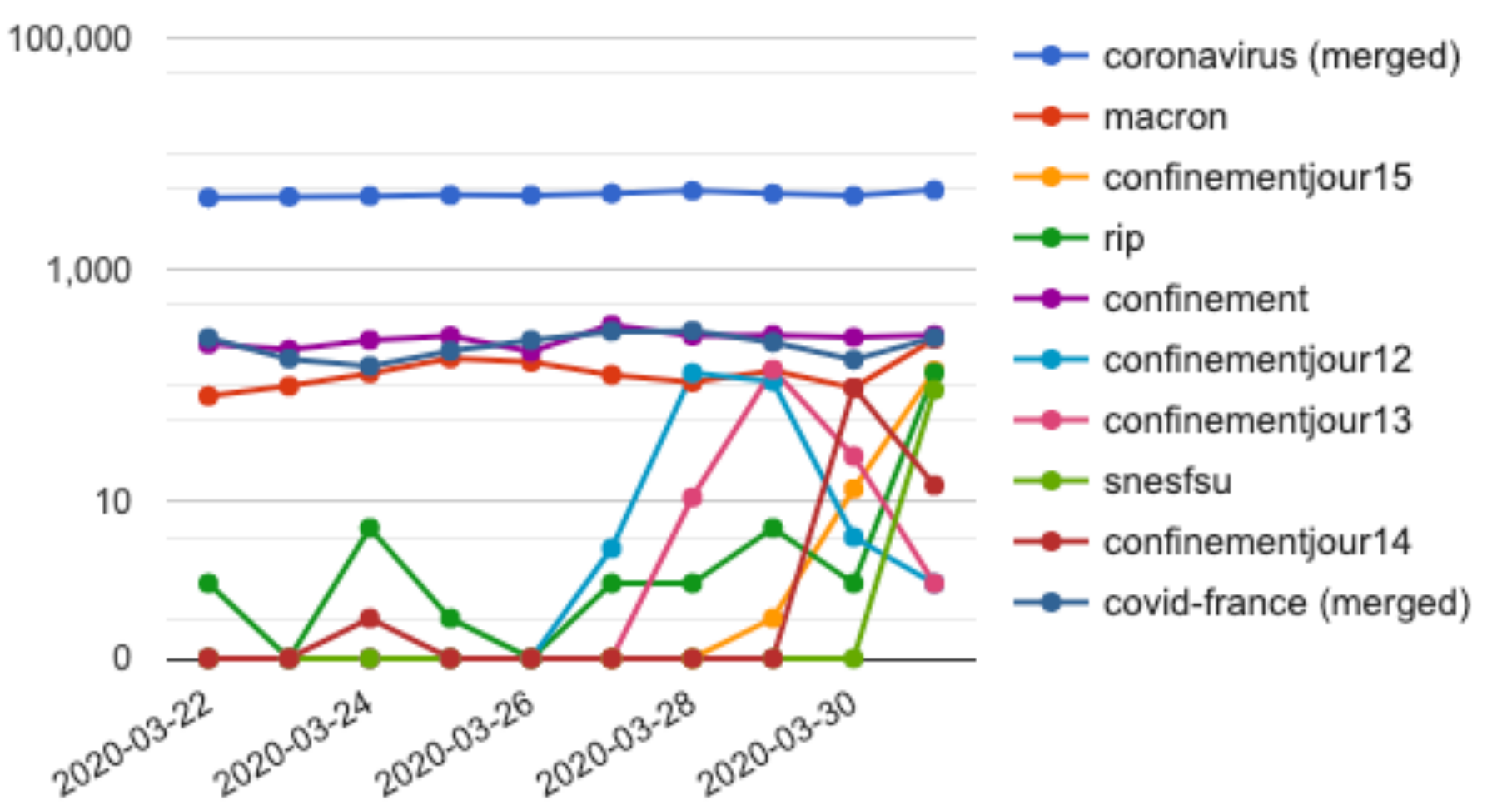}  
  \caption{France, March 22-31}
\end{subfigure}
\caption{Emerging hashtags for countries/regions for March 22-31, depicted for United states and France.}
\label{fig:dashboard-trend}
\end{figure}

\begin{figure}[t]

\begin{subfigure}{.5\textwidth}
  \centering
  \includegraphics[width=\linewidth,,height=4cm]{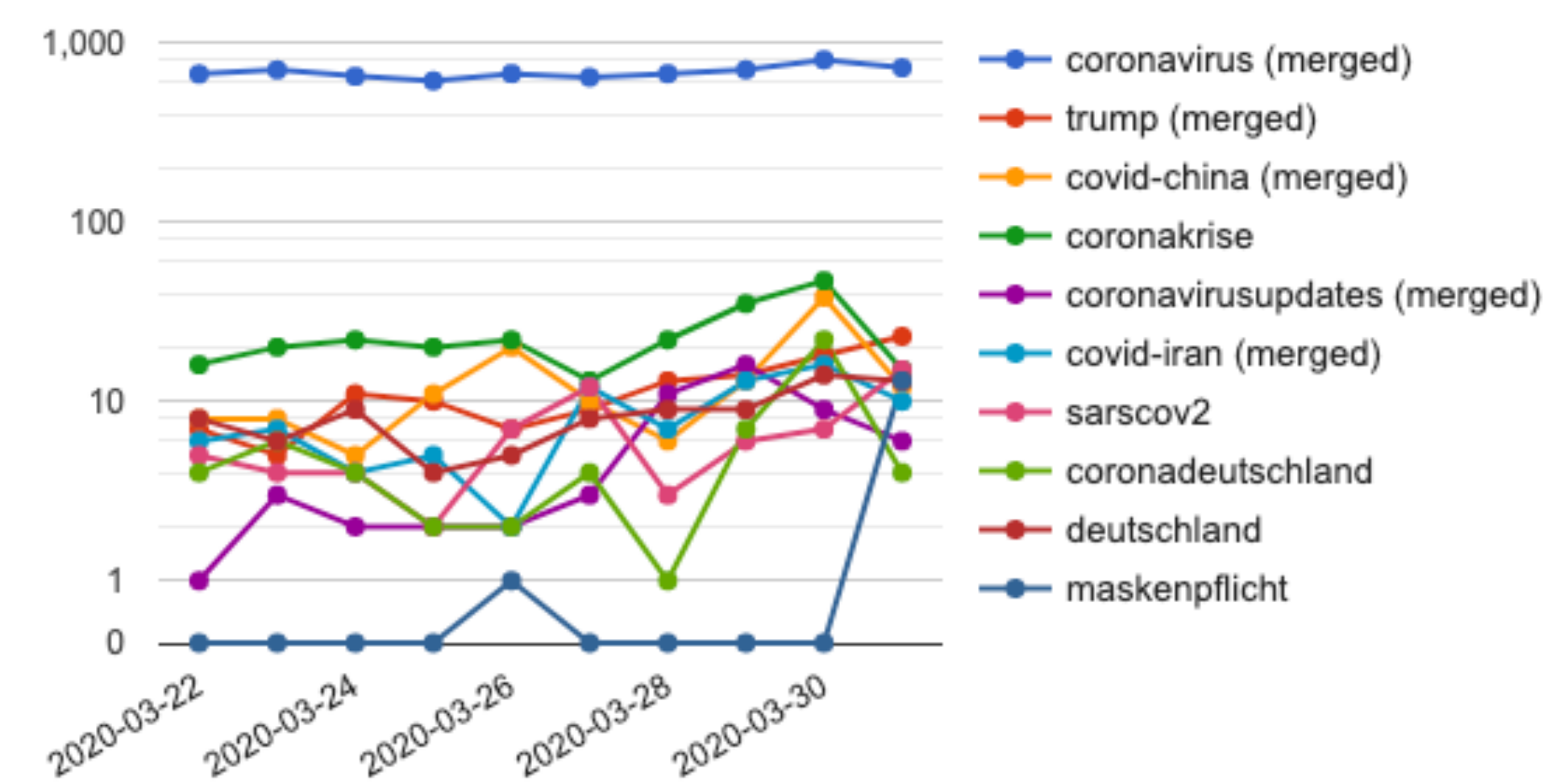}  
  \caption{Germany, March 22-31.}
\end{subfigure}
~
\begin{subfigure}{.5\textwidth}
  \centering
  \includegraphics[width=\linewidth,height=4cm]{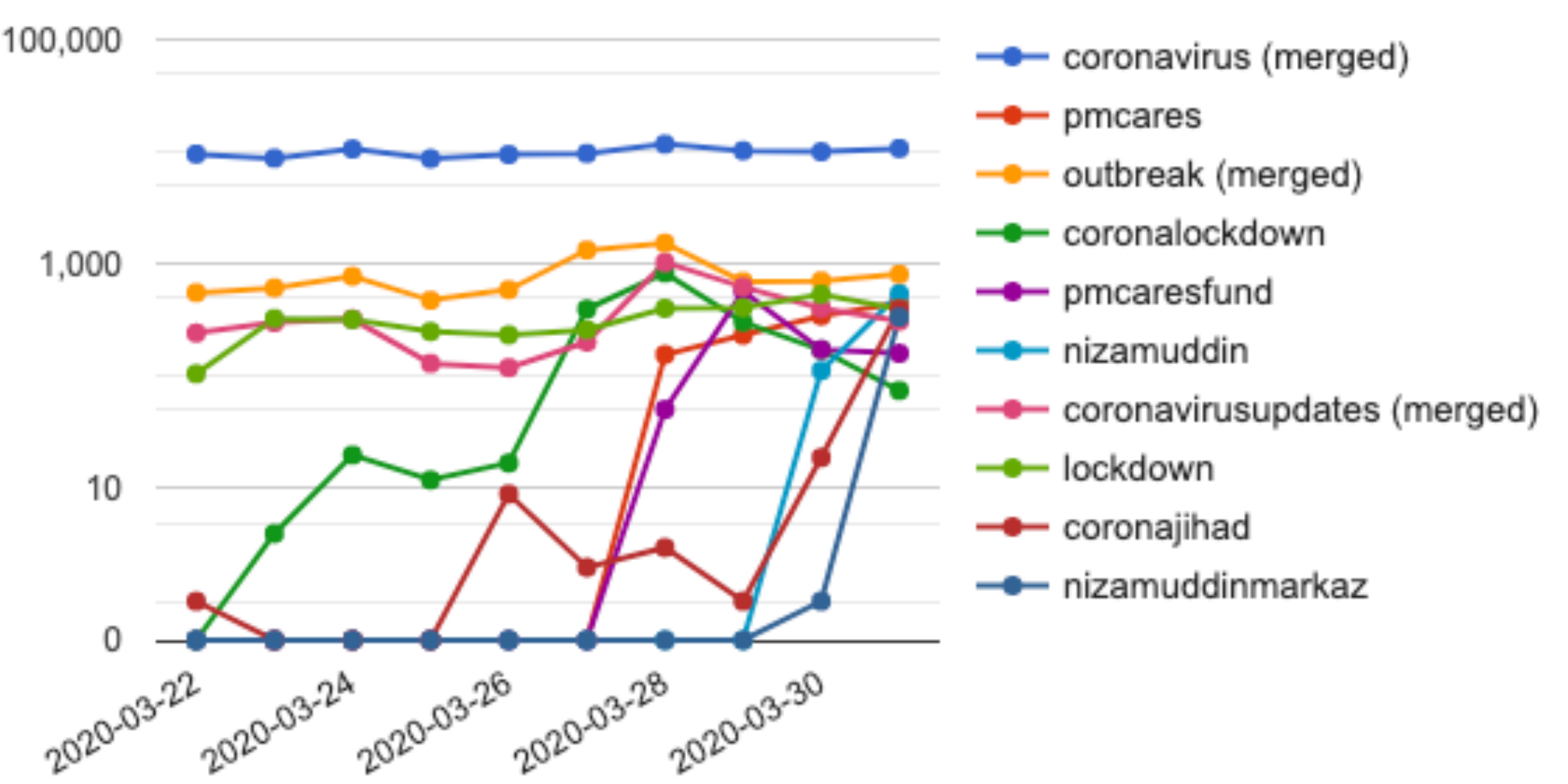}  
  \caption{India, March 22-31}
\end{subfigure}
\caption{Emerging hashtags for countries/regions for March 22-31, depicted for Germany and India.}
\label{fig:trend_countries}
\end{figure}

\textbf{Emerging trends.} The emerging trend on Twitter highlight changes in perception or importance of topics as the pandemic situation changes. 
We extract hashtags from the tweet text for all tweets in the dataset for March/2020.
The hashtags with emerging popularity are estimated based on fitted linear regression curves on the usage counts of hashtags over the period. 
On the dashboard (Fig.~\ref{fig:dashboard-trend}, Fig.~\ref{fig:trend_countries}), we provide the Top-30 emerging hashtags to show trendy interest in social media over the world.
As the hashtags also reflect spatial characteristics (e.g., country-level policy or trend), the Top-10 emerging hashtags of each country for last 10 days are also visualized on the dashboard and regularly updated.

The country/region-based emerging hashtags are particularly important to track people's interest.
For instance, the line chart in Fig.~\ref{fig:dashboard-trend} shows which hashtags emerged in terms of a slope of usage counts in the United States from March/22 to March/31. 
While the use of some hashtags (e.g., \texttt{\#coronavirus} and \texttt{\#trump}) continuously dominates the conversations, other hashtags (e.g., \texttt{\#coronalockdown}, \texttt{\#coronavirustruth} and \texttt{\#nationaldoctorsday}) are temporally significant.
The end of March is when most of states announced lockdown on many business and a stay-at-home order, and it causes people to use lockdown-related hashtags (\texttt{\#coronalockdown}).
Moreover, it shows that people get more and more interested in facts on coronavirus.
Finally, the slope-based extraction easily detects spike pattern of some hashtags (\texttt{\#nationaldoctorsday}), which are only used in a particular day.


In Germany (Fig.~\ref{fig:dashboard-trend}), we could detect that people are interested in wearing masks (\texttt{\#maskenpflicht}, mask required) from the end of March. In France (Fig.~\ref{fig:trend_countries})have counted the containment day (\texttt{\#confinementjour}) everyday and their patterns show time lags as expected.
Finally, the plot is also useful to see what trendy issues are (\texttt{\#coronajihad, \#nizamuddin}) in India (Fig.~\ref{fig:trend_countries}).

\textbf{Geoinformation trends.}
We also analyze the geographical distribution of daily counts of tweets and its trend using the extracted geolocation information. The dashboard provides (1) the geographical distribution of the daily count of tweets over countries/regions; (2) the daily increment of the count of tweets for each country/region; (3) the time for each country/region when it encounters its peak of daily counts of tweets. (1) shows a steady distribution of daily counts of tweets: users in United States contribute more than half of the total daily counts of tweets around the world, and users in Europe, India, Oceania and South America are also active. (2) reveals that the daily counts of tweets of most countries/regions are steady during the time of our observation. (3) illustrates the spatio-temporal pattern of which day each country/region achieves it highest activity over the observation time period.


\section{Conclusion and future work}
In this work, we provided analysis of social media discourse about COVID-19 on Twitter, through analysis of misinformation claims identified using information about low-quality news websites from fact-checking sources, and analysis of sentiments, topics, and emerging trends in the online discourse.
The dashboard presented analysis and daily updated list of identified misinformation claims between March to June, 2020 during the pandemic. There are several directions of future work to address this large-scale “infodemic" surrounding COVID-19. The
proportion of Twitter users in the United states is higher than in other countries like China with alternate social media
platforms. Since the pandemic is at a global scale, social media analysis for other platforms and languages is critical towards uncovering misinformation and tracking online discourse. Lastly, real-time tracking of misinformation, topics, trends, and sentiments, through public interfaces/applications (such as the dashboard) can enable solutions and provide an overview of the online discourse, to educate individuals on social media about the nature and quality of discussions on important topics. This can in turn make them less susceptible to misinformation, and prevent dire consequences of misinformation on social outcomes.


\bibliographystyle{plainnat}
\bibliography{main.bib}
\end{document}